%% file: paper.tex
\documentclass[sigplan,10pt]{acmart}

\usepackage[normalem]{ulem}

\acmYear{2026}\copyrightyear{2026}
\setcopyright{cc}
\setcctype[4.0]{by}
\acmConference[EUROSYS '26]{European Conference on Computer Systems}{April 27--30, 2026}{Edinburgh, Scotland UK}
\acmBooktitle{European Conference on Computer Systems (EUROSYS '26), April 27--30, 2026, Edinburgh, Scotland UK}
\acmDOI{10.1145/3767295.3769334}
\acmISBN{979-8-4007-2212-7/26/04}

\usepackage{booktabs}   
\usepackage{threeparttable}  
                        
\usepackage{listings}
\usepackage{algorithm}  
\usepackage{algorithmicx}
\usepackage{algpseudocode}  
\usepackage{subfig}

\usepackage{multirow}
\usepackage{tikz}
\usepackage[hyphenbreaks]{breakurl}
\usepackage{soul}
\usepackage{pifont}

\usepackage{bbding}

\lstdefinelanguage{java}{
  morekeywords={abstract,case,catch,class,def,%
    do,else,extends,false,final,finally,%
    for,if,implicit,import,match,mixin,%
    new,null,object,override,package, public,%
    private,protected,requires,return,sealed,%
    super,this,throw,trait,true,try,%
    type,int,double,while,pnew, persistable, atomic, durable\_root},
  otherkeywords={=>,<-,<\%,<:,>:,\#,@},
  sensitive=true,
  morecomment=[l]{//},
  morecomment=[n]{/*}{*/},
  morestring=[b]",
  morestring=[b]',
  morestring=[b]""",
}

\lstnewenvironment{q1}[1][]
{\lstset{
frame=none,
language=java,
aboveskip=3mm,
belowskip=3mm,
showstringspaces=false,
columns=flexible,
breaklines=true,
basicstyle={\footnotesize\ttfamily}, 
tabsize=4,  
numbers=left,
xleftmargin=2em,
framexleftmargin=2em,
#1} }
 {}
 
\begin{document}

\title{{\sys}: Protecting On-Device Large Language Models with Arm TrustZone}

\author{Xunjie Wang\textsuperscript{*}, Jiacheng Shi\textsuperscript{*}, Zihan Zhao, Yang Yu, Zhichao Hua, Jinyu Gu\textsuperscript{\Envelope}}
\affiliation{
\institution{Institute of Parallel and Distributed Systems, School of Computer Science, Shanghai Jiao Tong University}
\country{}
}

\renewcommand{\shortauthors}{X. Wang, et al.}

\thanks{\textsuperscript{*} The two authors contributed equally to this work.}

\newcommand{\sys}{TZ-LLM}
\newcommand{\us}{\textmu s}
\newcommand{\smc}{\textit{smc}}

\newcommand*\circled[1]{\tikz[baseline=(char.base)]{
            \node[shape=circle,draw,inner sep=0.1pt, minimum size=9pt] (char) {\footnotesize #1};}}

\newenvironment{myitemize}
  {\begin{list}{\labelitemi}{\itemsep1pt \topsep2pt \parsep0.00in
  \partopsep=0pt \leftmargin1.4em}}%
  {\end{list}}

\newcommand{\ra}[1]{\renewcommand{\arraystretch}{#1}}
\newcommand{\stitle}[1]{\vspace{0.5ex}\noindent{\bf #1}}

\newcommand{\pie}[1]{%
\begin{tikzpicture}
 \draw (0,0) circle (1ex);\fill[rotate=90] (1ex,0) arc (0:#1:1ex) -- (0,0) -- cycle;
\end{tikzpicture}%
}

\newcommand{\TODO}[1]{\textcolor{red}{TODO: #1}}

\date{}
\input{abs}



\maketitle
\thispagestyle{empty}


\input{intro}
\input{motivation}

\input{overview}

\input{design}
\input{impl}
\input{security-analysis}
\input{eval}
\input{related-work}
\input{conclusion}


\section*{Acknowledgments}
We sincerely thank our shepherd Hyungon Moon and the anonymous reviewers,
whose reviews, feedback, and suggestions have significantly strengthened our work.
This research was supported in part by
National Natural Science Foundation of China (No. 62432010),
CCF-Huawei Populus Grove Fund,
and the Fundamental Research Funds for the Central Universities.
Corresponding author: Jinyu Gu (\burl{gujinyu@sjtu.edu.cn}).


\bibliography{paper}




\input{appendix}

\end{document}

%% file: abs.tex
\begin{abstract}

Large Language Models (LLMs) deployed on mobile devices offer benefits like user privacy and reduced network latency, but introduce a significant security risk: the leakage of proprietary models to end users.

To mitigate this risk, we propose a system design for protecting on-device LLMs using Arm Trusted Execution Environment (TEE), TrustZone. Our system addresses two primary challenges: 
(1) The dilemma between memory efficiency and fast inference (caching model parameters within TEE memory).
(2) The lack of efficient and secure Neural Processing Unit (NPU) time-sharing between Rich Execution Environment (REE) and TEE.

Our approach incorporates two key innovations. First, we employ \emph{pipelined restoration}, leveraging the deterministic memory access patterns of LLM inference to prefetch parameters on demand, hiding memory allocation, I/O and decryption latency under computation time. Second, we introduce a \emph{co-driver} design, creating a minimal data plane NPU driver in the TEE that collaborates with the full-fledged REE driver. This reduces the TEE TCB size and eliminates control plane reinitialization overhead during NPU world switches.

We implemented our system on the emerging OpenHarmony OS and the llama.cpp inference framework, and evaluated it with various LLMs on an Arm Rockchip device. Compared to a strawman TEE baseline lacking our optimizations, our system reduces TTFT by up to 90.9\% and increases decoding speed by up to 23.2\%.

\end{abstract}

%% file: intro.tex
\section{Introduction}
\label{sec:intro}

Intelligent applications based on Large Language Models (LLMs), such as digital assistants, text refinement, and multi-modal understanding, have been increasingly deployed on mobile devices~\cite{apple-intelligence,samsung-galaxy-ai,huawei-celia}.
Compared to cloud-based LLMs, on-device LLMs can leverage the semantics of personal data without transmitting it to the cloud, which helps maintain user data privacy, reduce network latency, and lower the cost of LLM inference.
However, on-device LLMs introduce a new security challenge because the proprietary models are stored on personal devices.
A curious user or a malicious application may compromise the mobile OS to steal the models.
Although some model providers protect their models by encrypting the model files, the plaintext model in memory remains at risk during the inference process~\cite{mind-your-weights}.

Arm TrustZone~\cite{trustzone}, widely deployed on mobile devices, enforces isolation between a Rich Execution Environment (REE), which runs untrusted applications and a traditional OS, and a Trusted Execution Environment (TEE) for trusted applications (TAs) and a minimal TEE OS.
It is intuitive to run LLM inference in the TEE to protect the models.
However, traditional TEE software stacks, such as OP-TEE~\cite{op-tee}, are designed for lightweight TAs that typically require small computational resources.
How to efficiently provide the memory and Neural Processing Unit (NPU) resources for LLM inference in the TEE is challenging and lacks research.

This paper introduces {\sys}, a system designed to efficiently protect the confidentiality of on-device LLMs with TEE.
{\sys} overcomes the following two challenges.

The first challenge is a dilemma between memory efficiency and LLM startup time (time-to-first-token, TTFT).
On the one hand, if the TEE reserves a static amount of secure memory for caching LLM parameters,
the LLM inference can start immediately. However, the memory usage is inefficient due to the large size of LLM parameters (e.g., 8GB for 8-bit quantized Llama-3-8B). Mobile devices are resource constrained ($\leq$24GB memory for commodity smartphones).

On the other hand, scaling secure memory for the LLM on demand results in a long TTFT,
because the system needs to expand secure memory and load LLM parameters from flash storage before the LLM starts inference.
Due to the inherent requirement of TrustZone memory protection~\cite{tzc-400},
the system must allocate large contiguous physical memory from REE,
which is time-consuming~\cite{twinvisor,gcma,daac}.
Meanwhile, confidentiality protection requires model files to be encrypted, resulting in decryption overhead on model loading, in addition to I/O overhead.
The allocation, I/O and decryption overheads can increase the TTFT by 11.6s for Llama-3-8B.

The second challenge is the lack of an efficient and secure mechanism for NPU time-sharing between TEE and REE.
Existing work has shown that LLM performance can improve by up to {7.3\texttimes} with NPU~\cite{mllm,powerinfer-2},
while the applications in REE also require NPU for various functionalities~\cite{qnn-models}.
It is feasible to deploy two separated NPU drivers in REE and TEE and dynamically switch the NPU between the two worlds.
However, porting the REE NPU driver to TEE will lead to both performance and security concerns.
First, the REE driver and its dependencies on the REE OS consist of a large code base (e.g., 60K for the Rockchip NPU~\cite{rknpu-driver}), which can significantly bloat the TCB in TEE.
Second, when switching the NPU between the two worlds, the REE or TEE driver needs to be reinitialized, incurring a switching overhead (e.g., 32ms for the Rockchip NPU).

To address the first challenge, {\sys} proposes \emph{pipelined restoration} to overlap I/O, decryption, memory allocation, and inference, which allows it to dynamically scale secure memory while reducing the overhead on TTFT. 
The insight is that the memory usage pattern of the DAG-based (directed acyclic graph) inference computation is deterministic, allowing {\sys} to accurately prefetch parameters during inference.
Specifically, {\sys} incrementally extends the contiguous secure memory region with CMA~\cite{linux-cma} and loads parameters in the topological order of the DAG,
while executing inference operators in parallel when the required parameters are loaded.
To minimize pipeline bubbles, {\sys} uses the following two techniques.
\circled{1} \emph{Priority-based pipeline scheduling}: When multiple restoration or computation operators compete for the CPUs,
{\sys} prioritizes the most urgent task that could stall the critical path of the LLM inference.
\circled{2} \emph{Partial parameter caching}: If memory is sufficient when the LLM is idle,
{\sys} caches some parameters used by early prefill operators in the secure memory, allowing the next inference to start immediately.
The parameters are released in reverse topological order after inference, preserving the contiguity of the secure memory region.

To address the second challenge, {\sys} uses a \emph{co-driver} design that separates a data plane driver for TEE from the original NPU driver.
The insight is that the workflow of an NPU job is simple and can be secured without relying on the control plane,
so that the TEE driver can be tailored (about 1K LoC) and an NPU world switch does not require the reinitialization of the control plane.
The TEE driver is only responsible for launching secure NPU jobs, while the REE driver handles control plane operations like NPU job scheduling and device frequency management.
The two drivers cooperate: When the REE driver schedules a TEE NPU job, it asks the TEE driver to take over the NPU by configuring the TrustZone hardware and execute the job.

We prototype {\sys} based on OpenHarmony (a widely deployed production-grade mobile OS)~\cite{openharmony}
and the llama.cpp inference framework~\cite{llama-cpp},
and evaluate it with various models~\cite{tinyllama,qwen2_5,phi_3,llama_3} including Qwen2.5-3B and Llama-3-8B
and various real-world benchmarks~\cite{ultra_chat,personachat,autodroid} on an Orange Pi board~\cite{orange-pi} (Rockchip 3588 CPU/NPU~\cite{rk3588}).
We compare {\sys} to a strawman TEE baseline without pipelined restoration and NPU support, and to an REE baseline optimized with pipelined restoration.
Compared to the TEE baseline, {\sys} reduces TTFT by {76.1\%$\sim$90.9\%} and increases decoding speed by {0.9\%$\sim$23.2\%}.
Compared to the REE baseline, {\sys} incurs an average overhead of {5.2\%$\sim$28.3\%} and {1.3\%$\sim$4.9\%} on TTFT and decoding speed, respectively,
across different models and benchmarks.

In summary, this paper has the following contributions:
\begin{myitemize}
    \item We identify the challenges of efficient secure memory scaling and TEE-REE NPU time-sharing that hinders efficient LLM inference in TrustZone.

	\item We address these challenges and make the first attempt to use TrustZone to protect the confidentiality of on-device LLMs.
	
    \item Our preliminary evaluations show the promising performance of our system.

\end{myitemize}

%% file: motivation.tex
\section{Background and Motivation}
\label{sec:motivation}

\subsection{Confidentiality of On-Device LLMs}

\begin{figure*}[htbp]
    \centering
    \includegraphics[width=2.0\columnwidth]{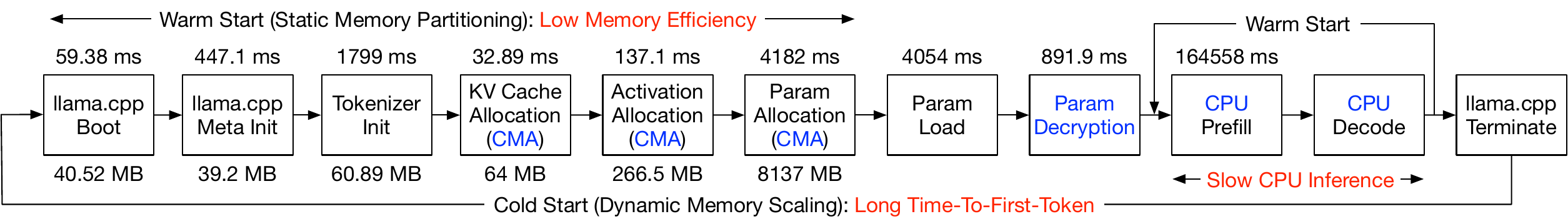}
    \caption{
      A strawman workflow of LLM inference in TEE (\textsection{\ref{sec:eval}} testbed, 8-bit Llama-3-8B, 512-token prompt).
      Time and memory usage for each step are shown above and below each box. 
      Red texts: challenges. Blue texts: overheads related to TEE protection.
    }
    \label{fig:inference-workflow}
\end{figure*}

Modern mobile applications can automatically incorporate user data into LLM prompts to generate personalized responses~\cite{apple-intelligence}.
Instead of using cloud systems, there is a growing trend to run LLM inference on mobile devices~\cite{powerinfer-2,llm-flash,mllm}, as it eliminates the network latency of querying cloud services and keeps users' private data on their devices.

However, storing LLM parameters on mobile devices introduces the risk of leaking the proprietary model to untrusted users, as mobile devices are prone to jailbreaking attacks.
Model leakage can result in significant financial losses for the model provider, as the development of such models may cost millions of dollars~\cite{mind-your-weights}. Additionally, the leakage could severely undermine the model provider's advantage in the highly competitive LLM market~\cite{lmarena}.

According to the prior study~\cite{mind-your-weights}, most mobile applications leave their on-device models completely unprotected, while others only encrypt model files but still allow attackers to extract plaintext model parameters from memory.

\subsection{Arm TrustZone}

Our work uses Arm TrustZone~\cite{trustzone}, a widely deployed hardware isolation mechanism on mobile devices, to protect LLMs.
TrustZone separates hardware resources into a Rich Execution Environment (REE) and a Trusted Execution Environment (TEE). 
The TEE hosts security-critical trusted applications (TAs) and a minimal TEE OS, while the REE runs untrusted applications and a full-fledged OS like Linux.

TrustZone divides the CPU into a secure state and a non-secure state.
Software can switch CPU states by calling a security monitor running in EL3 using a Secure Monitor Call ({\smc}) instruction.
To enforce memory isolation, a TrustZone Address Space Controller (TZASC) protects eight contiguous physical memory regions as secure memory, which cannot be accessed by non-secure CPUs.
Peripheral devices are also classified as secure devices and non-secure devices.
A TrustZone Protection Controller (TZPC) prohibits any MMIO access to secure devices from non-secure CPUs.
The TZASC controls the DMA permission of each device,
only allowing secure devices to access secure memory.
Moreover, TrustZone directs interrupts from secure devices to the TEE OS with an extension in the generic interrupt controller (GIC).

TrustZone can only protect contiguous physical memory,
but contiguous memory allocation at runtime is challenging due to fragmentation~\cite{mem-frag}.
Therefore, existing TEEs typically reserve secure memory at system boot.
The Linux kernel provides a Contiguous Memory Allocator (CMA)~\cite{linux-cma}, which reserves a physical memory region.
The buddy system can allocate pages from this region, but only \emph{movable} pages can be placed in it.
To preserve contiguity, CMA migrates \emph{movable} pages out of the region as follows:
the kernel allocates a new destination page outside CMA, unmaps the old page, copies its data to the new page, updates the page table mapping, and releases the old page for CMA allocation.

\subsection{Challenges of LLM Inference in TEE}
\label{sec:motivation-challenges}

\begin{table*}[htb]
  \fontsize{7pt}{7pt}\selectfont
  \centering
  \caption{Comparison of existing TEE-based model protection approaches with {\sys}.}
  \label{tab:existing-tee-approaches}
  \begin{threeparttable}
  \begin{tabular}{c|cc|c|cc|c}
    \toprule
    \multirow{2}{*}{\textbf{Approach}} & \multicolumn{2}{c|}{\textbf{Performance}} & \multirow{2}{*}{\textbf{End-to-end Security}} & \multicolumn{2}{c|}{\textbf{Compatibility}} & \multirow{2}{*}{\textbf{Memory Scaling}} \\

    & \textbf{Overall} & \textbf{Accelerator Usage} & & \textbf{No Model Modification} & \textbf{Quantization Support} \\
    \midrule
    Shielding the entire model\textsuperscript{1}    & \ding{72}                     & No              & \ding{52} & \ding{52} & \ding{52} & \ding{56} \\
    Obfuscation-based TSLP\textsuperscript{2}        & \ding{72} \ding{72}           & REE only        & \ding{56} & \ding{52} & \ding{56} & \ding{56} \\
    TSQP~\cite{sun2025tsqp}           & \ding{72} \ding{72}           & REE only        & \ding{56} & \ding{56} & \ding{52} & \ding{56} \\
    TEESlice~\cite{zhang2024no}       & \ding{72} \ding{72}           & REE only        & \ding{56} & \ding{56} & \ding{56} & \ding{56} \\
    StrongBox~\cite{strongbox}        & \ding{72} \ding{72}           & TEE-REE sharing & \ding{56} & \ding{52} & \ding{52} & \ding{56} \\
    SecDeep~\cite{secdeep}            & \ding{72} \ding{72}           & TEE only        & \ding{52} & \ding{52} & \ding{52} & \ding{56} \\
    {\sys} (ours)                     & \ding{72} \ding{72} \ding{72} & TEE-REE sharing & \ding{52} & \ding{52} & \ding{52} & \ding{52} \\
    \bottomrule
  \end{tabular}
  \begin{tablenotes}
    \item \textsuperscript{1} Shielding the entire model~\cite{yang2024penetralium,mlcapsule,occlumency,jian2025smartzone,privado,t-slice}.
          \textsuperscript{2} Obfuscation-based TSLP~\cite{wanggame,soter,zhang2024groupcover,li2024translinkguard,sun2023shadownet}.
  \end{tablenotes}
  \end{threeparttable}
\end{table*}

As illustrated in Figure~\ref{fig:inference-workflow}, running LLM efficiently in TEE faces the following two challenges.

\stitle{Challenge \#1: The dilemma between memory efficiency and fast inference.}
Traditional TEEs~\cite{op-tee} statically partition memory as secure and non-secure at system boot.
However, the LLM requires a large amount of memory for parameters, KV cache, activation, and other data (8.4GB in Figure~\ref{fig:inference-workflow}).
Using a large secure memory will result in memory shortage in REE as mobile devices are typically resource constrained.

Therefore, the secure memory should be dynamically scaled up and down as the LLM inference starts and completes.
However, when scaling up secure memory, a naive ``cold start'' workflow (Figure~\ref{fig:inference-workflow}) for restarting LLM Trusted Application (TA) will incur high overhead on LLM TTFT.
This overhead includes the following parts:
(1) The inference framework initializes, parses model metadata and creates the tokenizer (2.3s).
(2) The TEE allocates memory from REE. Due to the limitation of TZASC, it must allocate contiguous physical memory using Linux CMA,
causing high memory migration overhead if the CMA region is occupied (up to 4.2s for 8GB parameters).
(3) The system loads LLM parameters from the flash storage. Since the file system is accessible to the untrusted REE applications and OS, the model files must be encrypted, resulting in decryption overhead during loading (0.9s for 8GB parameters).
The total cold start overhead is 11.6s in Figure~\ref{fig:inference-workflow}.

Thus, a mechanism is needed to \emph{minimize the overhead on TTFT} caused by dynamic scaling of secure memory.

\stitle{Challenge \#2: The lack of efficient and secure NPU time-sharing between REE and TEE.}
NPUs are widely deployed on mobile devices to support applications such as object detection, OCR, and photo refinement~\cite{qnn-models}.
Since these applications typically run in the REE, the NPU is statically configured as a non-secure device at boot time.
However, this design significantly hinders LLM performance in the TEE.
As shown in Figure~\ref{fig:inference-workflow}, the LLM prefill using CPU takes 164s.
Prior work has shown that using the Qualcomm NPU~\cite{hexagon-npu} can increase LLM prefill speed by {7.3\texttimes} compared to the optimal CPU implementation~\cite{mllm}.
Our evaluation also shows that the Rockchip NPU provides {12.5\texttimes} and {1.3\texttimes} optimizations on the prefill and decoding speed of Llama-3-8B, respectively.

It is intuitive to share the NPU between REE and TEE by deploying one driver in each world.
The NPU can be switched between the two worlds by detaching it from one driver and attaching it to another driver.
However, this approach has two limitations:
(1) The detach-attach incurs substantial switching overhead as it requires full driver reinitialization.
The detach-attach of a Rockchip NPU with the Linux driver takes 32ms.
The overhead mainly stems from control plane operations, including NPU power/frequency configuration and interaction with the Linux device framework.
(2) Deploying the full-fledged NPU driver, which highly depends on the REE OS, in the TEE bloats the TCB.
The Linux driver for Rockchip NPU relies on several Linux subsystems, like device, memory, interrupt, and power management,
and the total code base is estimated to be over 60K LoC.

Thus, a mechanism for NPU time-sharing between REE and TEE is needed to \emph{reduce NPU world switching overhead} and \emph{minimize the additional TCB in TEE}.

\subsection{Existing Approaches Studies}
\label{sec:motivation-existing-approaches-studies}

\subsubsection{TEE-based Model Protection}
\label{sec:motivation-tee-based-model-protection}

As shown in Table~\ref{tab:existing-tee-approaches},
extensive prior work has explored protecting on-device models with TEEs such as Arm TrustZone and Intel SGX.
Some work~\cite{yang2024penetralium,mlcapsule,occlumency,jian2025smartzone,privado,t-slice}
shields the entire model within an accelerator-absent TEE
to protect all model parameters and the inference framework.
Although these approaches offer end-to-end security guarantees, they only use CPU for inference and incur significant overhead.
Consequently, a line of work seeks to mitigate this overhead.

\stitle{TEE-Shielded LLM Partition (TSLP).}
TSLP solutions partition models and offload a part of the parameters to REE accelerators for computation.
Some TSLP approaches~\cite{wanggame,soter,zhang2024groupcover,li2024translinkguard,sun2023shadownet}
enhance security by obfuscating the offloaded parameters.
However, TSQP~\cite{sun2025tsqp} points out that these approaches are incompatible with quantization,
while quantization significantly reduces memory footprint and is well-suited for mobile devices~\cite{on-device-llm,survey-resource-efficient-llm}.
TSQP enables quantization through quantization-aware model training.
Nonetheless, model stealing attacks using public pretrained models can compromise the security of these obfuscation-based TSLP solutions~\cite{zhang2024no}.
TEESlice~\cite{zhang2024no} counters this threat with privacy-aware model training, offloading only privacy-irrelevant parameters to REE accelerators.
However, it requires model modification, and it still leaves part of the parameters outside the TEE, failing to provide end-to-end security guarantees.
In addition, these solutions also incur extra data copying between the REE and TEE,
as well as additional computation for deobfuscation or privacy-related parameters.

\stitle{Accelerator-enabled TEE.}
Other work attempts to enable accelerators within the TEE.
StrongBox~\cite{strongbox} builds a GPU TEE with TrustZone by deploying the a single GPU driver in the REE for both secure and non-secure jobs.
While it supports page-grained secure memory protection with S2PT, it reserves a static secure memory region and resorts to TZASC for DMA protection,
which is not memory efficient.
Moreover, it does not safeguard the integrity of the inference framework in the REE, thus lacking end-to-end security guarantees.
SecDeep~\cite{secdeep} statically configures accelerators as TrustZone secure devices, which restricts the REE functionalities.
These approaches also incur frequent encryption and decryption overhead when data is swapped between secure and non-secure memory during inference.

In conclusion, existing TEE-based model protection systems fail to meet performance, security, and compatibility requirements simultaneously, primarily because they lacks REE-TEE accelerator time-sharing or dynamic secure memory scaling for effcient model inference inside the TEE.

\subsubsection{Elastic Secure Memory Protection}

Previous work uses Stage-2 Page Tables (S2PT) for memory protection at page granularity~\cite{strongbox,blackbox,mytee}.
Specifically, they run the untrusted OS and applications inside a VM and unmap the secure memory pages from the S2PT.
Although this design could support elastic secure memory scaling without the overhead of contiguous memory allocation,
we conduct preliminary experiments on the testbed in \textsection{\ref{sec:eval}} to demonstrate why we choose not to adopt it.

\begin{figure}[htb]
  \centering
  \includegraphics[width=1\columnwidth]{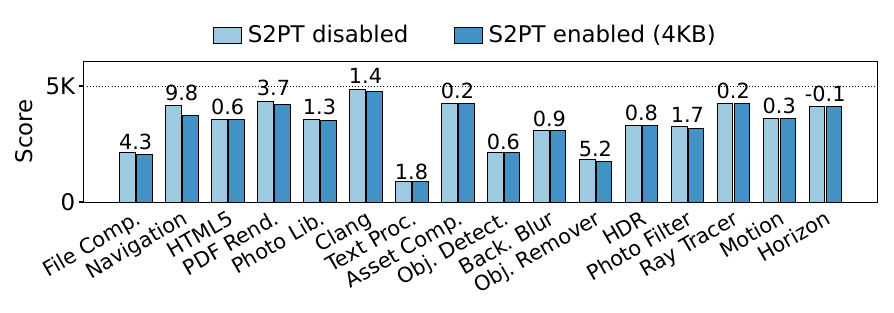}
  \caption{Geekbench scores with S2PT enabled or disabled. The texts are the overheads caused by S2PT (\%).}
  \label{fig:meta-geekbench}
\end{figure}

\stitle{S2PT incurs continuous overhead on REE applications, while the overhead of CMA allocation is transient.}
The CPU running REE applications must perform a two-dimensional page table walk for each TLB miss~\cite{dmt,addr-passthrough,agile-paging}.
Although using 2MB or 1GB huge pages in the S2PT reduces overhead,
most mappings fall back to 4KB granularity after allocating memory for the LLM due to memory fragmentation.
Figure~\ref{fig:meta-geekbench} shows that stage-2 translation with 4KB mappings can incur a maximum overhead of 9.8\% on Geekbench applications~\cite{geekbench}, and the average overhead is 2.0\%.

Although stage-2 translation can be disabled to avoid the overhead when the LLM is idle,
this disables memory protection, requiring all secure memory to be cleaned.
To mitigate model loading overhead, parameters can be cached in S2PT-protected memory, at the cost of \emph{continuous} overhead on REE applications.

In contrast, while migrating pages from the CMA region also imposes overhead on REE applications,
the overhead is \emph{transient} and only exists at the beginning of LLM inference.

\begin{figure}[htb]
  \centering
  \includegraphics[width=1\columnwidth]{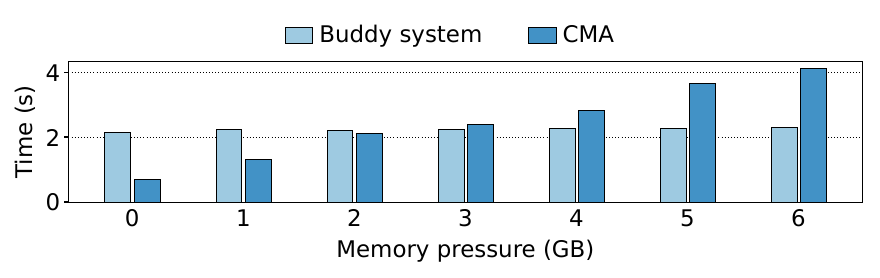}
  \caption{Memory allocation time for Llama-3-8B (8GB) using buddy system or CMA, at different memory pressures.}
  \label{fig:meta_mem_alloc}
\end{figure}

\stitle{CMA allocation overhead is small under low memory pressure, and can be hidden under high memory pressure.}
We evaluate the time for allocating 8GB memory for 8-bit Llama-3-8B using CMA and buddy system (4KB), respectively.
To assess the overhead of page migration, we use stress-ng~\cite{stress-ng},
which maps a portion of memory and generates pressure by running multiple sophisticated memory testing algorithms on the mapped region.
Figure~\ref{fig:meta_mem_alloc} shows the results.

Under high memory pressure, the CMA allocation throughput is 1.9GB/s, which is similar to the I/O throughput of sequential reads on our platform (2GB/s).
Moreover, by using multi-threading, the CMA allocation throughput can reach 3.8GB/s (4 threads).
Therefore, we can hide the allocation overhead under the latency of reading the model file.

\stitle{S2PT protection cannot prevent DMA attacks.}
S2PT does not control DMA permissions.
To prevent DMA attacks on S2PT-protected secure memory, a privileged monitor like the EL3 monitor must intercept every IOMMU configuration operation and unmap the secure memory from I/O page tables~\cite{blackbox}, or intercept every MMIO operation and verify the DMA addresses~\cite{mytee}.
Both designs introduce monitoring overhead on REE and extend the privileged TCB.

%% file: overview.tex
\section{Overview}
\label{sec:overview}

\subsection{Threat Model}

\stitle{Attack vectors.}
We consider an attacker attempting to steal on-device LLM parameters, or intermediate inference results that may help model theft, such as activations and KV cache.
The attacker might extract parameters by directly accessing memory/flash or exploiting peripheral devices to initiate malicious DMA requests.
Attackers may also try to induce the inference framework to exfiltrate model parameters, by exploiting TEE-REE interfaces for Iago attacks (e.g., breaking the integrity of secure NPU jobs).
Physical attacks on memory confidentiality are not considered because TrustZone does not enforce memory encryption, and it can be addressed with future hardware~\cite{arm-cca}.
Side-channel and cryptographic attacks fall outside our scope as they are orthogonal to our concern and can be defended with complementary techniques.
Denial-of-service (DoS) attacks are also out-of-scope as they do not compromise model confidentiality.

\stitle{Trusted computing base (TCB).}
We trust the TEE OS, the TEE NPU driver, and the inference framework (LLM TA).
Arm TrustZone hardware, EL3 monitor, and NPU hardware are also trusted.
The integrity of these components can be guaranteed with secure boot.
Other components within the TEE, such as other TAs and other secure devices, are not trusted.
All components in the REE are excluded from the TCB,
including the REE OS, the full-fledged REE NPU driver, REE applications, and non-secure peripheral devices.

\subsection{System Architecture}
\label{sec:system-architecture}

\begin{figure}[htbp]
    \centering
    \includegraphics[width=1.0\columnwidth]{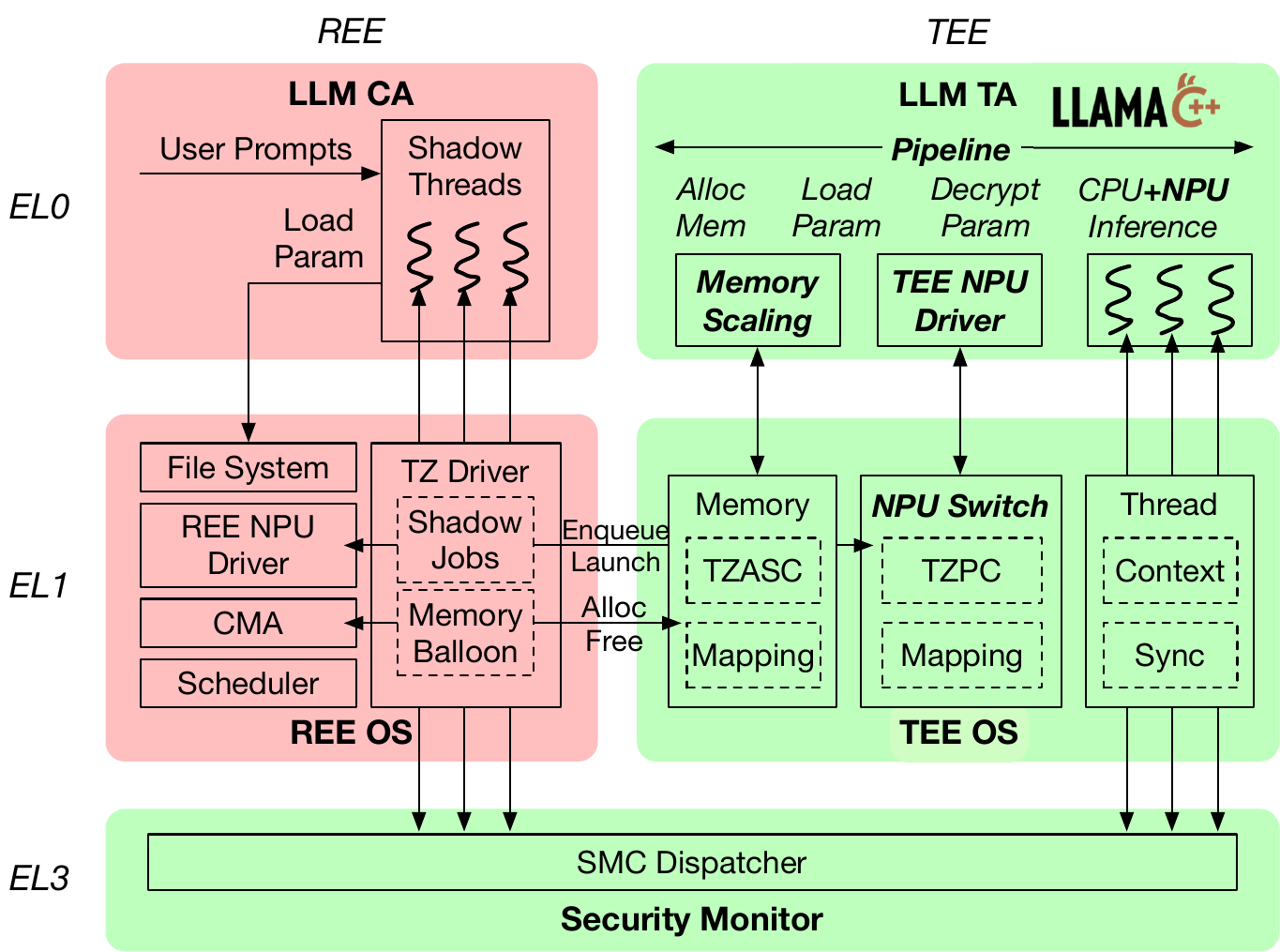}
    \caption{{\sys} architecture, S/N: secure/non-secure.}
    \label{fig:architecture}
\end{figure}

We propose {\sys}, a system for protecting on-device LLMs using Arm TrustZone.
As shown in Figure~\ref{fig:architecture},
{\sys} runs the LLM inference framework (e.g. llama.cpp) as a TA,
which can be invoked by a client application (CA) in the REE
through the TrustZone (TZ) driver in the REE OS (Linux).
The TZ driver also enables interactions between the TEE OS and the CA, Linux CMA, and REE NPU driver to delegate model loading, memory scaling, and NPU job scheduling.

Our design assumes a mobile platform with a hardware platform supporting Arm TrustZone and a software platform consisting of a REE OS with a TEE OS.
These assumptions are generally applicable across mobile devices.

\stitle{Addressing challenge \#1: Elastic memory scaling with pipelined restoration.}
The LLM TA can extend or release secure memory using interfaces provided by the TEE OS.
When extending the secure memory, the TEE OS asks the TZ driver to allocate memory from Linux CMA (memory ballooning) and protects it by configuring TZASC.
The extended memory can be released with a reverse process.

To mitigate the parameter restoration overhead (allocation, I/O, and decryption) on LLM TTFT (Figure~\ref{fig:inference-workflow}), {\sys} runs these processes in parallel with LLM inference.
The insight for this design is the \emph{determinism} of the memory access pattern of LLMs.
Specifically, the computation graph of the LLM is a DAG, in which each node represents an operator like matrix multiplication, and the inference framework schedules the operator in the topological order of the DAG.
Each operator only uses a portion of LLM parameters, e.g., operators in LLM layer 1 only use parameters of layer 1.
Therefore, when handling one operator, the inference framework can accurately know which parameters will be accessed next and prefetch these parameters in parallel.

If an LLM operator is ready, but the parameters have not been restored, or the hardware is busy, the operator will be blocked, leading to pipeline bubbles.
{\sys} designs two techniques to minimize such bubbles (\textsection{\ref{sec:pipeline}}).
First, the pipeline is scheduled using a priority-based and preemptive mechanism that prioritizes the most urgent task in the pipeline that may lead to a bubble.
Second, the LLM TA uses a partial caching mechanism that gradually releases memory based on the REE memory pressure after the inference is done, and the parameters remaining in memory can be used by the next inference without restoration.

With partial parameter caching, the TA must ensure that the cached secure memory is contiguous.
Fortunately, it is optimal to cache the parameters used early during inference, so that the secure memory is released in the reverse topological order of the DAG.
This first-in-last-out allocation-deallocation pattern aligns well with the contiguity requirement.
We design an \emph{``extend and shrink''} secure memory management interface based on this pattern (\textsection{\ref{sec:memory-scale}}).

\stitle{Addressing challenge \#2: TEE-REE NPU time-sharing with control-data separation.}
The LLM TA can issue secure NPU jobs with a TEE NPU driver.
The TEE and the REE multiplex the NPU with time-sharing.
An REE application can run NPU jobs during LLM inference.

For secure and efficient TEE-REE NPU time-sharing,
we observe that the workflow of an NPU job (data plane), including setup, launching, and completion, forms a small and self-contained closure.
The functionality and security of this workflow does not depend on the control plane state of the full-fledged NPU driver, such as scheduling or power management.
This property allows {\sys} to use a \emph{co-driver} design (\textsection{\ref{sec:npu-sharing}}) by integrating only the tiny data plane of the NPU driver into the TEE, which cooperates with the control plane in the REE NPU driver.
Therefore, most control plane code and dependencies can be tailored from the TEE driver and the NPU can switch between the two worlds without reinitializing the control plane.
The REE driver manages the unified scheduling of both secure and non-secure NPU jobs, delegating secure jobs to the TEE driver.
The TEE driver protects the confidentiality and integrity of the secure jobs based on TrustZone hardware configuration and security checks.

\stitle{Other techniques for efficient inference.}
As shown in Figure~\ref{fig:inference-workflow}, the initialization of framework, model metadata and tokenizer also takes a long time.
We mitigate this overhead by saving a checkpoint of the initialized state in flash and restoring it on each inference request.
The KV cache and activation allocation overhead is not mitigated because it is minor compared with the inference time.

In addition to NPU, on-device LLM inference also requires CPU multi-threading for acceleration,
but traditional TEEs provide only one thread for each TA.
{\sys} allows the TA to create multiple threads and schedules them using the REE scheduler.
Specifically, each TA thread is paired with a shadow thread in the CA.
When a shadow thread is activated, it uses {\smc} to start or resume the corresponding TA thread.
For security, the contexts of TA threads and the synchronization primitives are managed by the TEE OS.

The file system is managed by the REE, and the LLM TA delegates I/O requests to the CA with {\smc} when loading parameters from the flash.
To avoid blocking the CPU, the CA issues asynchronous I/O (aio) requests to the file system.

%% file: design.tex
\section{Detailed Design}
\label{sec:design}

\subsection{Pipelined Parameter Restoration}
\label{sec:pipeline}

\begin{figure*}[htbp]
    \centering
    \includegraphics[width=2.0\columnwidth]{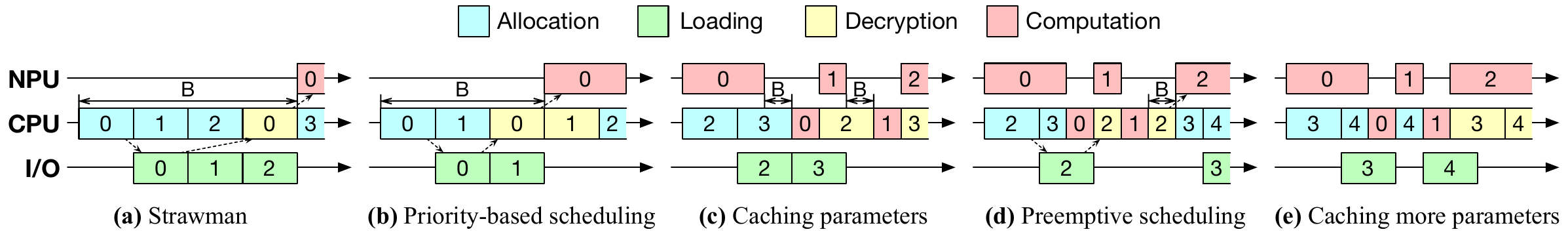}
    \caption{
        Pipelined restoration timelines.
        The figure shows the effect of different techniques for reducing bubbles.
        B: bubble.
        The number in each box denotes the index of the computation operator that the operator belongs to.
        The indices follow the topological order of the computation graph.
        The dashed arrows denote the dependencies of operators, which cause bubbles.
    }
    \label{fig:pipeline-timeline}
\end{figure*}

To accelerate the cold start of LLM TA, {\sys} adopts a pipeline mechanism that overlaps the parameter restoration operations with the prefill-stage computation of the LLM.

\begin{figure}[htbp]
    \centering
    \includegraphics[width=1.0\columnwidth]{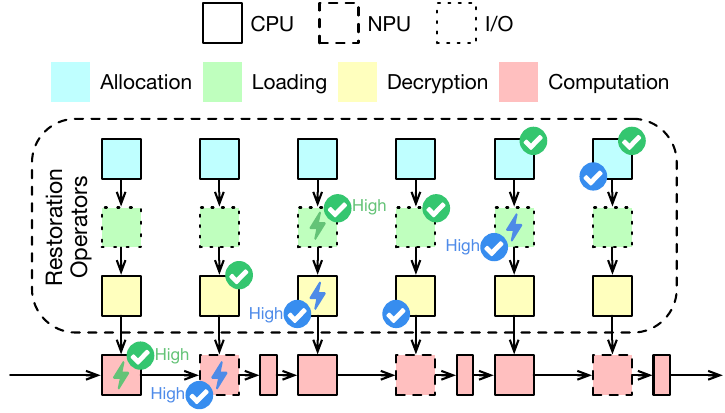}
    \caption{
        Pipeline scheduling examples.
        The arrows denote the dependencies of operators.
        The green and blue marks denote two different scheduling points.
        The checkmark denotes that the operator is ready.
        The lightning symbol denotes that the operator is scheduled (highest priority).
    }
    \label{fig:pipeline-schedule}
\end{figure}

\stitle{Restoration operators.}
As shown in Figure~\ref{fig:pipeline-schedule}, with parameter restoration, the LLM computation graph is extended by inserting three restoration operators before a prefill-stage computation operator, representing the memory allocation, parameter loading (flash I/O), and parameter decryption for restoring the parameters used by the computation operator.

The computation and restoration operators run on three types of hardware: CPU, NPU, and I/O engine.
Contiguous memory allocation (memory migration) and parameter decryption run on CPUs.
Some computation operators, such as layer normalization and self-attention, run on CPUs, while others, such as matrix multiplication, run on the NPU.
Parameter loading is performed by the I/O engine.

\stitle{Pipeline scheduling problem.}
There may be multiple restoration or computation operators ready for the same hardware at the same time.
For example, in Figure~\ref{fig:pipeline-schedule}, the green checkmarks indicate that there are four operators ready for the CPUs and two operators ready for the I/O engine.
The scheduling problem is to determine the execution order of the operators to minimize the TTFT.
For the scheduling of the I/O engine and the NPU, it is intuitive that the best policy is to schedule the loading (I/O) and computation operators in the topological order of the computation graph.
Therefore, the main problem is the scheduling of CPU operators.

\stitle{Priority-based pipeline scheduling.}
It is hard to find an optimal scheduling policy for CPU operators because the scheduling goal depends on the pipeline critical path, which varies across different models, prompts, and hardware.
There are three potential critical paths:
(1) loading (I/O) operators,
(2) CPU operators, including allocation, decryption, and computation,
and (3) computation operators, including CPU and NPU computation.
If loading operators are the critical path, the scheduler should prioritize allocation operators to reduce bubbles on the loading path.
If computation operators are the critical path, the scheduler should prioritize computation operators and make restoration operators complete early enough to prevent computation stalls.
If CPU operators are the critical path, the scheduler should keep the CPU busy, reducing bubbles caused by waiting for I/O or NPU.

In practice, we observe that the critical path is usually CPU operators or computation operators, instead of loading operators.
To meet the scheduling goals of these two common critical paths,
{\sys} uses a greedy policy that schedules the CPU computation operator if it's ready, or schedules the restoration operator related to the earliest computation operator if no CPU computation operator is ready.
This aligns well with the two scheduling goals because
(1) it reduces computation stalls by prioritizing computation operators and earlier restoration operators, and
(2) it enables CPU computation operators to be ready for scheduling early, thereby keeping the CPU busy.
The evaluation shows that the performance of this policy is close to the optimal (\textsection{\ref{sec:scheduling-policy-effectiveness}}).

The scheduler maintains a priority queue of ready operators and executes them according to the priority rule.
As shown in Figure~\ref{fig:pipeline-timeline}a and Figure~\ref{fig:pipeline-timeline}b, the scheduler prioritizes decryption operator 0 over allocation operator 2, reducing the bubble before NPU computation operator 0.

\stitle{Preemptive pipeline scheduling.}
We find that priority-based scheduling without preemption still suffers from bubbles due to the misalignment of operator execution times.
As shown in Figure~\ref{fig:pipeline-timeline}c, CPU computation operator 0 is blocked by allocation operator 3, resulting in a bubble.
We eliminate such bubbles with preemptive scheduling, by dividing allocation and decryption operators into smaller micro-operators and introducing preemption points between them.
As shown in Figure~\ref{fig:pipeline-timeline}d, allocation operator 3 is preempted as soon as CPU computation operator 0 becomes ready.

\stitle{Partial parameter caching.}
As shown in Figure~\ref{fig:pipeline-timeline}b, the pipeline has a bubble at the beginning that waits for the first parameter tensor, regardless of the scheduling policy.
To eliminate this bubble, {\sys} keeps some secure memory unrevoked after inference to partially cache the plaintext parameters.
With this mechanism, the next inference can resume from the computation stage of the cached parameters, avoiding full restoration.
As shown in Figure~\ref{fig:pipeline-timeline}c, by caching parameters of operators 0$\sim$1, the initial bubble is eliminated.

Sometimes {\sys} needs to cache more parameters, as the bottleneck of the prefill stage may shift to restoration operators when the computation time is short.
For example, computation operator 2 in Figure~\ref{fig:pipeline-timeline}d is blocked by decryption operator 2, and this bubble can be eliminated by caching parameters of operators 0$\sim$2 (Figure~\ref{fig:pipeline-timeline}e).

It is optimal to cache the parameters used by early computation operators as the later parameters can be restored in parallel with the early computation.
To this end, {\sys} lazily releases secure memory in the reverse topological order of the computation graph according to the REE memory pressure.
The LLM TA provides an interface to the REE OS to revoke secure memory to the REE.

\stitle{Limitation.}
A limitation of {\sys} is that it may deliver suboptimal performance on non-deterministic workloads.
For example, it prefetches all experts in a Mixture of Experts (MoE) model or all layers in an early-exit transformer,
including parameters not used in the current inference.
The cost of this additional prefetching can be amortized by future inferences that do utilize these parameters.

\subsection{Pipeline-Aware Secure Memory Management}
\label{sec:memory-scale}

The limitation of TZASC mandates that secure memory remain contiguous during scaling.
Fortunately, the memory allocation-deallocation pattern of pipelined restoration allows us to design secure memory management interfaces that effectively satisfy this requirement.

\stitle{Allocation patterns and memory layouts.}
The LLM TA uses four types of data: LLM parameters, KV cache, activations, and others (libraries, metadata, etc.).
{\sys} places these data in two contiguous TZASC regions.

\begin{figure}[htbp]
    \centering
    \includegraphics[width=1.0\columnwidth]{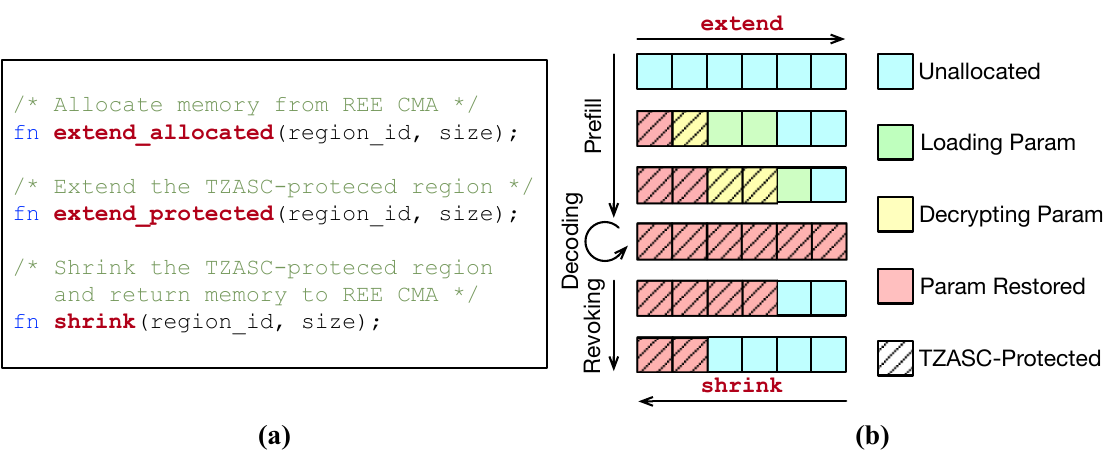}
    \caption{
        (a) Secure memory management interfaces,
        (b) memory layout of the CMA region for model parameters.
        }
    \label{fig:memory-interface}
\end{figure}

One TZASC region is used for LLM parameters.
With partial parameter caching (\textsection{\ref{sec:pipeline}}),
LLM parameters are progressively loaded during pipelined restoration and progressively released in the reverse order of allocation after inference.
As shown in Figure~\ref{fig:memory-interface}b, this first-in-last-out allocation-deallocation pattern ensures that the in-memory parameters are always stored contiguously.

Another TZASC region is used for KV cache, activations, and other data.
The KV cache is initialized to the prompt size during the prefill stage,
grows with the number of generated tokens during the decoding stage,
and is completely released after inference.
The activations and other data are fixed-size buffers allocated at inference start and released at inference completion,
so that they can be placed before the KV cache without breaking the contiguity of the TZASC region.

\stitle{Secure memory management interfaces.}
Based on the allocation patterns and memory layouts, the TEE OS provides \emph{``extend and shrink''} interfaces to the LLM TA for scaling TZASC regions up and down, as shown in Figure~\ref{fig:memory-interface}a.

Each TZASC region is associated with a CMA region in the REE.
When extending the secure memory, the TA first calls \emph{extend\_allocated}.
Then, the TEE OS asks the TZ driver to allocate memory blocks from the CMA region.
To ensure the contiguity of the entire allocated memory, CMA allocates new memory blocks adjacent to the previously allocated blocks.
The TEE OS verifies this requirement when it receives the allocated memory address from the TZ driver.
After allocation, the TA calls \emph{extend\_protected}.
The TEE OS then extends the end of the TZASC region to protect the newly allocated memory, and maps the new memory into the TA's address space.
When revoking secure memory, the TA calls \emph{shrink} to release memory from the end of the TZASC region.
The TEE OS unmaps memory from the TA's address space, shrinks the TZASC region and asks the TZ driver to release memory to the CMA.
The TEE OS clears all sensitive data before releasing the memory.

The separation of \emph{extend\_allocated} and \emph{extend\_protected} is designed to eliminate the need for I/O bounce buffers during parameter loading (flash I/O).
As shown in Figure~\ref{fig:memory-interface}b, after calling \emph{extend\_allocated},
the REE file system can directly load encrypted parameters into the \emph{unprotected} allocated memory, instead of a bounce buffer.
After loading, the new memory is protected with \emph{extend\_protected} and the parameters are decrypted.
This design reduces memory consumption and avoids additional copying overhead.

\stitle{Minimizing TEE OS modification.}
The ``extend and shrink'' interfaces introduce only minor modifications to the TEE OS.
In contrast, if the LLM TA is allowed to allocate/deallocate secure memory in an arbitrary order,
the TZASC region will become fragmented and the TEE OS needs to defragment the region upon revocation.
{\sys} leverages the allocation patterns to avoid this complexity.

\subsection{TEE-REE NPU Time-Sharing}
\label{sec:npu-sharing}

\begin{figure}[htbp]
    \centering
    \includegraphics[width=0.9\columnwidth]{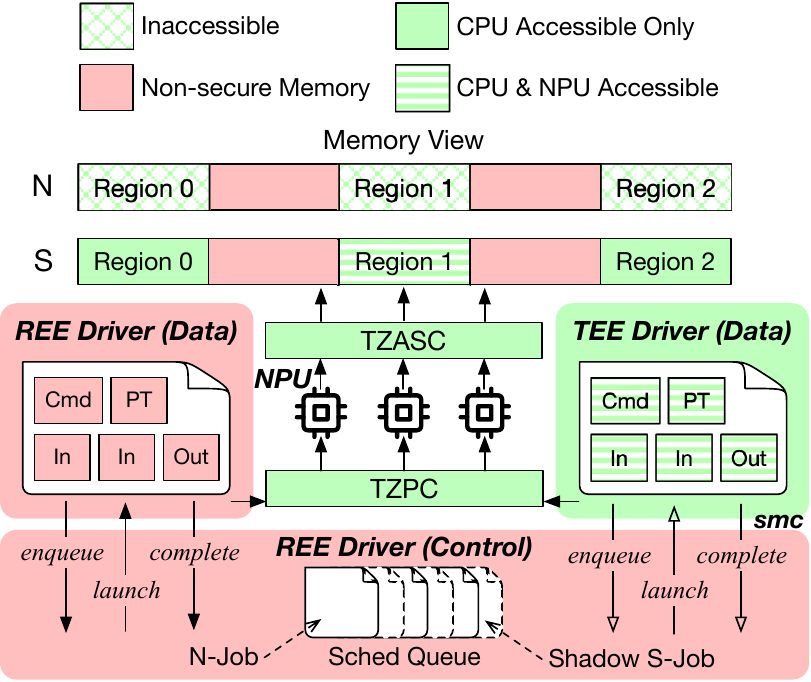}
    \caption{TEE-REE NPU time-sharing, S: secure, N: non-secure, cmd: register commands, PT: I/O page table, in/out: input/output buffers, region: TZASC region.}
    \label{fig:npu-task-protection}
\end{figure}

Inspired by the outsource-and-verify principle~\cite{zhou2014dancing},
which delegates complex operations to an untrusted component while verifying their outcomes,
{\sys} adopts a co-driver design to enable NPU time-sharing between the REE and the TEE,
as shown in Figure~\ref{fig:npu-task-protection}.
Specifically, a full-fledged NPU driver and a small data plane NPU driver are deployed in the REE and the TEE, respectively, cooperating to manage secure and non-secure NPU jobs.
The TEE data plane driver outsources control plane operations to the untrusted REE driver and verifies the returned results.

The goals of the co-driver design are to:
(1) separate the NPU driver into isolated domains,
(2) provide an isolated execution environment that ensures the confidentiality and integrity of secure NPU jobs,
(3) enable NPU time sharing for secure and non-secure NPU jobs with minimal overhead, and
(4) minimize the additional TCB introduced to the TEE.

\stitle{Separating control and data planes.}
The data plane of the NPU driver performs the following steps for each NPU job:
(1) it initializes the execution context of the job, i.e., the \emph{memory} for the I/O page table, register commands (the NPU job code), and input/output buffers;
(2) it performs \emph{MMIO} operations to launch the job by specifying the execution context;
(3) it handles \emph{interrupts} upon job completion.
These steps form a minimal closure that should be integrated into the TEE driver,
with the corresponding resources (\emph{memory}, \emph{MMIO}, and \emph{interrupts}) isolated to preserve the confidentiality and integrity of secure NPU jobs.

The control plane of the NPU driver manages device configuration during initialization and power management before and after job execution.
As shown in Figure~\ref{fig:npu-task-protection}, it also handles job scheduling, which interacts with the data plane through scheduling interfaces:
(1) it \emph{enqueues} the job into the scheduling queue;
(2) it calls the data plane for \emph{launching} when the job is scheduled;
(3) it continues to schedule the next job upon \emph{completion} of the current job.
Since the control plane does not access the isolated resources during job execution,
it can safely reside in the REE driver.
The function call interfaces between the REE control plane and the TEE data plane are replaced with {\smc}.

\stitle{Isolated execution environment.}
The TEE driver switches the NPU between non-secure and secure modes.
In non-secure mode, the NPU is prohibited from accessing secure memory.
In secure mode, the NPU's \emph{MMIO} region is accessible only to the TEE, its \emph{interrupts} are routed to the TEE, and it is allowed to access secure \emph{memory}.
Secure jobs run in secure mode, and the execution contexts of secure jobs are stored in secure memory.

Specifically, the TEE driver performs the following steps when switching the NPU to secure mode.
First, it updates the TZPC to isolate the MMIO region of the NPU from the REE and the GIC controller to route NPU interrupts to the TEE.
Second, it waits for the ongoing non-secure NPU job, if any, to complete.
Third, it sets the TZASC to grant the NPU access to secure memory.
The order of these steps is critical to ensure that
(1) no new non-secure NPU job can be launched during the sanity check of ongoing non-secure jobs, and
(2) any previously launched non-secure NPU job is completed before the NPU is granted access to secure memory. 

\stitle{TEE-REE time-sharing.}
{\sys} reuses the NPU job scheduling mechanism in the REE driver to support NPU time-sharing for secure and non-secure NPU jobs.

As shown in Figure~\ref{fig:npu-task-protection}, the REE driver is extended to maintain a unified scheduling queue for secure and non-secure NPU jobs.
Each time the LLM TA issues a secure NPU job, the TEE driver issues a paired shadow job with an empty execution context to the REE driver.
When a shadow job is scheduled, the REE driver proactively transfers NPU control to the TEE driver.
The TEE driver then transitions the NPU into secure mode to create an isolated execution environment.
To prevent arbitrary launch and replay attacks, the TEE driver ensures that the secure NPU job has been previously initialized but not yet issued.
To prevent reordering attacks, the TEE driver assigns each job a monotonic sequence number before issuing it to the REE driver
and verifies the number against the current execution sequence number when scheduled.
After these checks, the TEE driver launches the secure NPU job and waits for its completion.
Upon completion of the secure job (receipt of a secure interrupt), the TEE driver returns the NPU back to non-secure mode and informs the REE driver that the shadow job is complete.
Finally, the REE driver discards the shadow job and schedules the next NPU job.

\stitle{Minimal TCB.}
Despite the complexity of the REE NPU driver, {\sys} minimizes the additional TCB in TEE with two complementary approaches.
First, {\sys} integrates only the tiny data plane closure into the TEE driver,
while excluding control plane components such as job scheduling and dependencies on complex Linux subsystems like device, memory, interrupt, and power management.

Second, {\sys} deprivileges the TEE NPU driver to user mode,
isolating potential vulnerabilities in the driver from affecting the existing TEE system.
The TEE OS strictly confines the privileges of the user-mode NPU driver by enforcing two restrictions.
First, the TEE OS only maps the MMIO region of the NPU into the NPU driver's address space,
and thus the driver cannot access other secure devices.
Second, the TEE OS only allows the NPU to access the execution contexts of secure NPU jobs.
This is possible because the parameters, intermediate results, I/O page tables and register commands are placed in independent TZASC regions (\textsection{\ref{sec:memory-scale}}).
By configuring the TZASC, the TEE OS only allows the NPU to access these specific regions, while prohibiting NPU access to all other regions.
This design follows the broader minimal TCB TEE philosophy~\cite{costan2016sanctum,lee2020keystone},
retaining only a minimal privileged security monitor for isolation while deprivileging functionality into user mode.

%% file: impl.tex
\section{Implementation}
\label{sec:impl}

We prototype {\sys} on OpenHarmony OS~\cite{openharmony} and its TEE system,
which is an open-sourced version of Huawei's commercial HarmonyOS~\cite{harmonyos}.
The LLM TA is built based on llama.cpp~\cite{llama-cpp}, a popular on-device inference framework.
The REE OS is OpenHarmony v4.1 with Linux v5.10. The NPU driver is Rockchip NPU driver v0.9.8~\cite{rknpu-driver}.

The original TEE OS contains 17K LoC for basic functionalities, including thread management, IPC, interrupt dispatching, and memory management.
We only extend it with 62 LoC to manage CMA page memory mapping and 50 LoC to support dynamic configuration of TZASC and TZPC.
The llama.cpp inference framework is extended with
1.2K LoC for pipelined restoration, 1K LoC for integrating the data plane of the NPU drive, and the OpenSSL library~\cite{openssl} for parameter decryption.
Note that the computation graph is directly extracted via internal interfaces of llama.cpp.
In the REE OS, we add only 364 LoC to the Linux kernel, which consists of 167 LoC in the NPU driver for shadow job scheduling and 197 LoC in the TZ driver for CMA allocation and deallocation.

The current implementation works on a Rockchip platform, while {\sys} design is applicable to other Arm platforms, such as Qualcomm.
We investigate the open-source Linux driver for Qualcomm NPUs~\cite{qnn-driver} and confirm that we can also extract a small data plane driver from it.

%% file: security-analysis.tex
\section{Security Analysis}
\label{sec:security-analysis}

{\sys} protects the confidentiality of LLM parameters from any attacker who compromise the REE OS, the REE applications, or other TAs in the TEE.

\stitle{Preventing direct access attacks.}
If an attacker in the REE tries to access plaintext parameters in the secure memory, the TZASC hardware blocks such access attempts.
A malicious TA also cannot access the parameters in secure memory as the TEE OS enforces address space isolation between TAs.

If the attacker attempts to read the parameters in flash, he/she will only get content encrypted with a model key.
The model key in flash is encrypted with a hardware-protected TEE key.
It can only be decrypted by the TEE OS.
The TEE OS only allows the LLM TA to access the model key.

\stitle{Preventing DMA attacks.}
The attacker may exploit the NPU or other untrusted devices to initiate malicious DMA requests targeting parameters in the secure memory.

For the NPU, the TEE driver enforces two key protections before granting access to secure memory.
First, it configures the TZPC to prohibit REE access to MMIO interface of the NPU.
Second, it ensures that no NPU job previously launched by the REE driver is still executing.
Therefore, the DMA destination can only be a benign address set by the TEE driver, preventing parameter leakage.

For untrusted devices, whether secure or non-secure, the TZASC is configured to reject any access from them to the secure memory regions for LLM parameters.

\stitle{Preventing Iago attacks.}
Attackers may attempt to compromise the LLM TA or TEE OS for model theft by exploiting the interface between the TEE and the REE for Iago attacks.
{\sys} exposes four TEE-REE interfaces vulnerable to Iago attacks: secure memory scaling, NPU job scheduling, model loading, and CPU thread scheduling.

For secure memory scaling, the CMA may return arbitrary memory addresses to the TEE.
{\sys} counters this by validating the contiguity of the returned address against the previously allocated memory (\textsection{\ref{sec:memory-scale}}).
For NPU job scheduling, the REE NPU driver may schedule unauthorized secure jobs, replay previously scheduled jobs, or reorder them.
{\sys} counters this by validating the job before execution (\textsection{\ref{sec:npu-sharing}}).
For model loading (\textsection{\ref{sec:system-architecture}}), a malicious REE OS may return forged results.
{\sys} counters this by verifying the returned content using checksums.
For CPU thread scheduling, the REE scheduler may violate the required execution order of TA threads.
{\sys} counters this by managing synchronization primitives in the TEE (\textsection{\ref{sec:system-architecture}}),
ensuring that TA thread follows the execution order enforced by these primitives.

\stitle{Side-channel and physical attack considerations.}
Existing side-channel attacks on TrustZone~\cite{Alias-Driven,TruSpy,ARMageddon} are outside the scope of this paper and have known mitigations~\cite{HybCache,DAWG}.
{\sys} may introduce two other side channels.
First, the parameter tensor sizes are exposed to the REE when the TA scales secure memory.
Second, the execution time of secure NPU jobs is exposed to the REE driver when it schedules the jobs.
These channels may reveal model structures, but not parameter values.
To the best of our knowledge, there are no public reports of side-channel attacks successfully stealing on-device LLM parameters.
Additionally, these channels could be mitigated through orthogonal techniques such as dummy parameter loading and dummy computation.

Physical attacks through offline DRAM analysis, such as cold-boot attacks~\cite{cold-boot}, stem from TrustZone's hardware limitations.
They can be mitigated by future memory encryption hardware~\cite{arm-cca} and are orthogonal to {\sys}.

\stitle{Ensuring the security of the existing TEE system.}
{\sys} minimizes its security impact on the existing TAs and TEE OS.
First, {\sys} minimizes the modification of the privileged TEE OS to about 100 LoC (\textsection{\ref{sec:impl}}), by running the TEE NPU driver in user space, and designing simple secure memory scaling interfaces (\textsection{\ref{sec:memory-scale}}).
Second, even if the LLM TA is compromised, it cannot access the memory of other TAs or the TEE OS with direct read/write or NPU DMA,
because the TEE OS enforces address space isolation, and the TZASC configuration only allows the NPU to access the memory regions for NPU job execution contexts (\textsection{\ref{sec:npu-sharing}}).

%% file: eval.tex
\section{Performance Evaluation}
\label{sec:eval}

In this section, we first evaluate the end-to-end performance of the LLM TA (\textsection{\ref{sec:end2end}}) and analyze the optimization effect of pipelined restoration (\textsection{\ref{sec:pipelined-restoration}}).
Then, we run the LLM TA in parallel with REE applications to evaluate the overhead of TEE-REE NPU time-sharing (\textsection{\ref{sec:npu-multiplexing}}) and the interference caused by CMA allocation (\textsection{\ref{sec:normal-app-overhead}}).

\stitle{Testbed.}
The evaluation is conducted on an Orange Pi 5 Plus board~\cite{orange-pi} (RK3588 CPU/NPU~\cite{rk3588}),
equipped with four Cortex-A76 (2.4GHz) CPU cores and four Cortex-A55 (1.8GHz) CPU cores, 16GB of LPDDR4X RAM, an 1TB NVMe SSD (PCIe 3.0 x4), and an NPU with three cores and up to 6 TOPS of computation power.

\stitle{Baselines.}
We compare {\sys} with the following baselines:
(1) \emph{REE-LLM-Memory}: The unmodified llama.cpp framework running in the REE, with all model parameters preloaded in memory.
This baseline represents the theoretical best performance,
but it is impractical due to memory inefficiency and lacks protection for parameters.
(2) \emph{REE-LLM-Flash}: The unmodified llama.cpp framework running in the REE,
with model parameters loaded with pipelined restoration at inference start (buddy system allocation, no decryption).
This baseline is practical but provides no protection for parameters.
(3) \emph{Strawman}: The \emph{``cold start''} strawman in \textsection{\ref{sec:motivation-challenges}}, which performs cold start and CPU computation for each inference request.
This baseline offers security guarantees and memory efficiency but lacks pipelined restoration and NPU support within the TEE.

\stitle{Models and deployment.}
We evaluate {\sys} with four representative on-device LLMs:
TinyLlama-1.1B\cite{tinyllama},
Qwen2.5-3B\cite{qwen2_5},
Phi-3-3.8B\cite{phi_3},
and Llama-3-8B\cite{llama_3}.
All models are 8-bit quantized, with parameter sizes of 1.0 GB, 3.3 GB, 3.7 GB, and 7.9 GB, respectively.

The LLM TA runs on the four Cortex-A76 CPU cores and all three NPU cores.
For evaluations in \textsection{\ref{sec:end2end}} and \textsection{\ref{sec:pipelined-restoration}}, we simulate the memory pressure in the REE with the stress-ng~\cite{stress-ng} tool, to trigger memory migration during CMA allocation. 
To show the worst-case performance, the memory pressure is 13GB, 11GB, 10GB and 6GB for the four models, respectively.
The stressing threads and LLM threads are pinned to different CPU cores to avoid interference.

\stitle{Benchmarks.}
We use three benchmarks from prior work on on-device LLMs~\cite{mllm,powerinfer-2}:
UltraChat~\cite{ultra_chat} (multi-turn dialogues),
PersonaChat~\cite{personachat} (chat summarization tasks),
and DroidTask~\cite{autodroid} (UI automation tasks).

\subsection{End-to-End Performance}
\label{sec:end2end}

In this section, we evaluate the prefill and decoding performance of the LLM TA and explain the source of overhead or optimization compared with the baselines.

\subsubsection{Prefill Performance}
\label{sec:eval-prefill-performance}

We evaluate {\sys}'s end-to-end prefill performance using both fixed-length prompts and real-world benchmarks.

\begin{figure}[htbp]
    \centering
    \includegraphics[width=1.0\columnwidth]{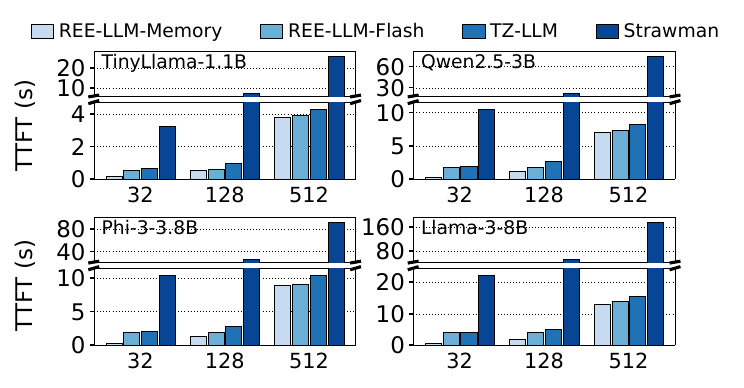}
    \caption{TTFT of different models under different prompt lengths. The x-axis represents the prompt length.}
    \label{fig:end2end}
\end{figure}

\stitle{TTFT under different prompt lengths.}
Figure~\ref{fig:end2end} presents the TTFT of the evaluated systems and models at prompt lengths of 32, 128 and 512 tokens.

Compared to the strawman baseline, {\sys} reduces the TTFT by 77.1\%$\sim$91.1\% across all models and prompt lengths.
This improvement stems from the pipelined parameter restoration mechanism, the NPU support in the TEE, and the checkpoint/restoration of the framework initial state (mentioned in \textsection{\ref{sec:overview}}).
The NPU support accelerates prefill computation, reducing TTFT by up to 87.2\%.
Meanwhile, state checkpoint eliminates the framework initialization overhead, further reducing TTFT by up to 36.8\%.
Finally, the pipeline mechanism effectively hides the parameter restoration latency, further reducing TTFT by up to 40.6\%.

Compared to the REE-LLM-Flash baseline, {\sys} incurs 2.5\%$\sim$22.3\%, 22.2\%$\sim$55.3\%, 10.2\%$\sim$15.2\% overhead at prompt lengths of 32, 128, and 512, respectively.
This overhead is mainly caused by CMA allocation (memory migration) and the parameter decryption during parameter restoration.
The overhead is relatively small for short and long prompt lengths, but more pronounced for medium prompt lengths.
For short prompt lengths, the TTFT is bounded by flash I/O, allowing allocation and decryption to overlap with I/O operators.
For long prompt lengths, the TTFT is dominated by CPU and NPU computation, enabling allocation and decryption to overlap with NPU execution.
In contrast, for medium prompt lengths, the TTFT is driven by the sum of CPU computation, allocation, and decryption,
resulting in only partial overlap of allocation and decryption with NPU execution and I/O operators.
Other overheads of {\sys}, such as the communication between TEE/REE NPU drivers and decryption of the framework initial state, are minor compared with prefill computation time.

{\sys} incurs up to {8.5\texttimes} overhead compared with the REE-LLM-Memory baseline, due to parameter restoration.
However, {\sys} is more memory-efficient, and the overhead can be reduced with partial parameter caching (\textsection{\ref{sec:cached-llm-states}}).
Moreover, the overhead is only 13.0\%$\sim$18.9\% for the long prompts (512 tokens), as parameter restoration overhead is hidden under the computation time.

\begin{figure}[htbp]
    \centering
    \includegraphics[width=1.0\columnwidth]{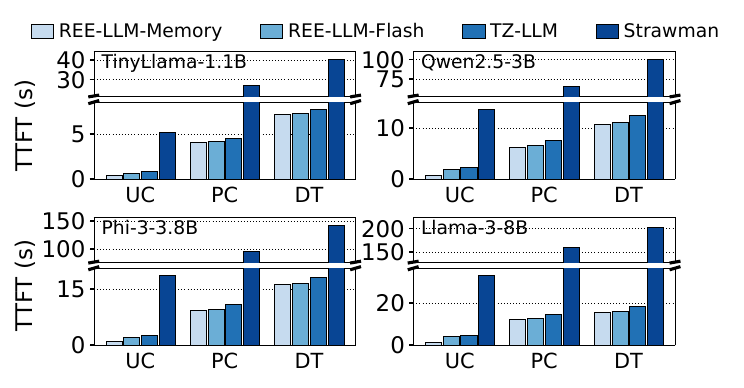}
    \caption{Average TTFT on different real-world benchmarks, UC: UltraChat, PC: PersonaChat, DT: DroidTask.}
    \label{fig:end2end_prefill_benchmark}
\end{figure}

\stitle{Benchmark results.}
Figure~\ref{fig:end2end_prefill_benchmark} shows the average TTFT of the evaluated systems and models on the three benchmarks.

For each pair of model and benchmark, we calculate the geometric mean of {\sys}'s overhead/optimization across different prompts.
{\sys} achieves 76.1\%$\sim$90.9\% TTFT reduction compared to the strawman baseline, while incurring 5.2\%$\sim$28.3\% slowdown compared to the REE-LLM-Flash baseline.
Compared to the REE-LLM-Memory baseline, {\sys} incurs {2.5\texttimes}$\sim${3.7\texttimes} overhead on UltraChat and {8.1\%}$\sim${21.2\%} overhead on PersonaChat and DroidTask.
The higher overhead on UltraChat is due to its shorter prompts,
where parameter restoration dominates the inference time,
but it can be mitigated via partial parameter caching (\textsection{\ref{sec:cached-llm-states}}).

\subsubsection{Decoding Performance}
\label{sec:eval-decoding-performance}

\begin{figure}[htbp]
    \centering
    \includegraphics[width=1.0\columnwidth]{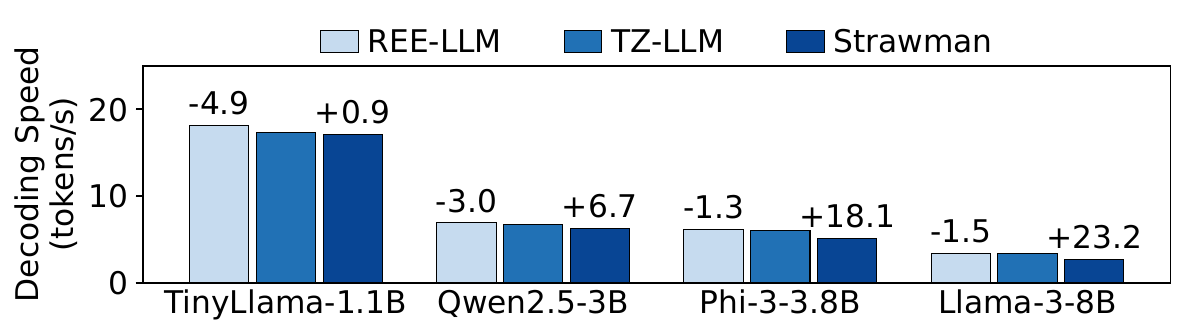}
    \caption{
        Token generation speeds during decoding for different models.
        The percentages shown above each bar represent {\sys}'s relative performance improvement (+) or degradation (-) compared to the respective baseline.
    }
    \label{fig:end2end_decoding}
\end{figure}

Figure~\ref{fig:end2end_decoding} shows the decoding speed at a prompt length of 128 and an output length of 64.
Results under other prompt and output lengths are similar and are omitted for brevity.
The decoding speeds of REE-LLM-Memory and REE-LLM-Flash are the same, so we only show a single bar.

The decoding speed of {\sys} shows a modest 0.9\%$\sim$23.2\% improvement over the strawman baseline, thanks to the NPU support in the TEE.
In contrast to the more significant gains seen in the prefill stage,
this relatively small improvement can be attributed to the single-batch computation pattern of decoding (processing one token in each iteration),
which cannot fully utilize the computation power of the NPU.
Compared to the REE-LLM baseline, {\sys} experiences a 1.3\%$\sim$4.9\% slowdown in decoding speed.
This overhead originates from the communication between the TEE and REE NPU drivers for NPU multiplexing (\textsection{\ref{sec:npu-sharing}}).
The overhead is smaller for larger models because the NPU computation time is longer.

\subsection{Effect of Pipelined Restoration}
\label{sec:pipelined-restoration}

In this section, we comprehensively evaluate the pipelined restoration mechanism in {\sys}.
First, we evaluate the effectiveness of our pipeline scheduling policy (\textsection{\ref{sec:scheduling-policy-effectiveness}}).
Then, we analyze how preemptive scheduling (\textsection{\ref{sec:pipeline-breakdown}}) and partial parameter caching (\textsection{\ref{sec:cached-llm-states}}) reduce the TTFT.

\begin{figure}[htbp]
    \centering
    \includegraphics[width=1.0\columnwidth]{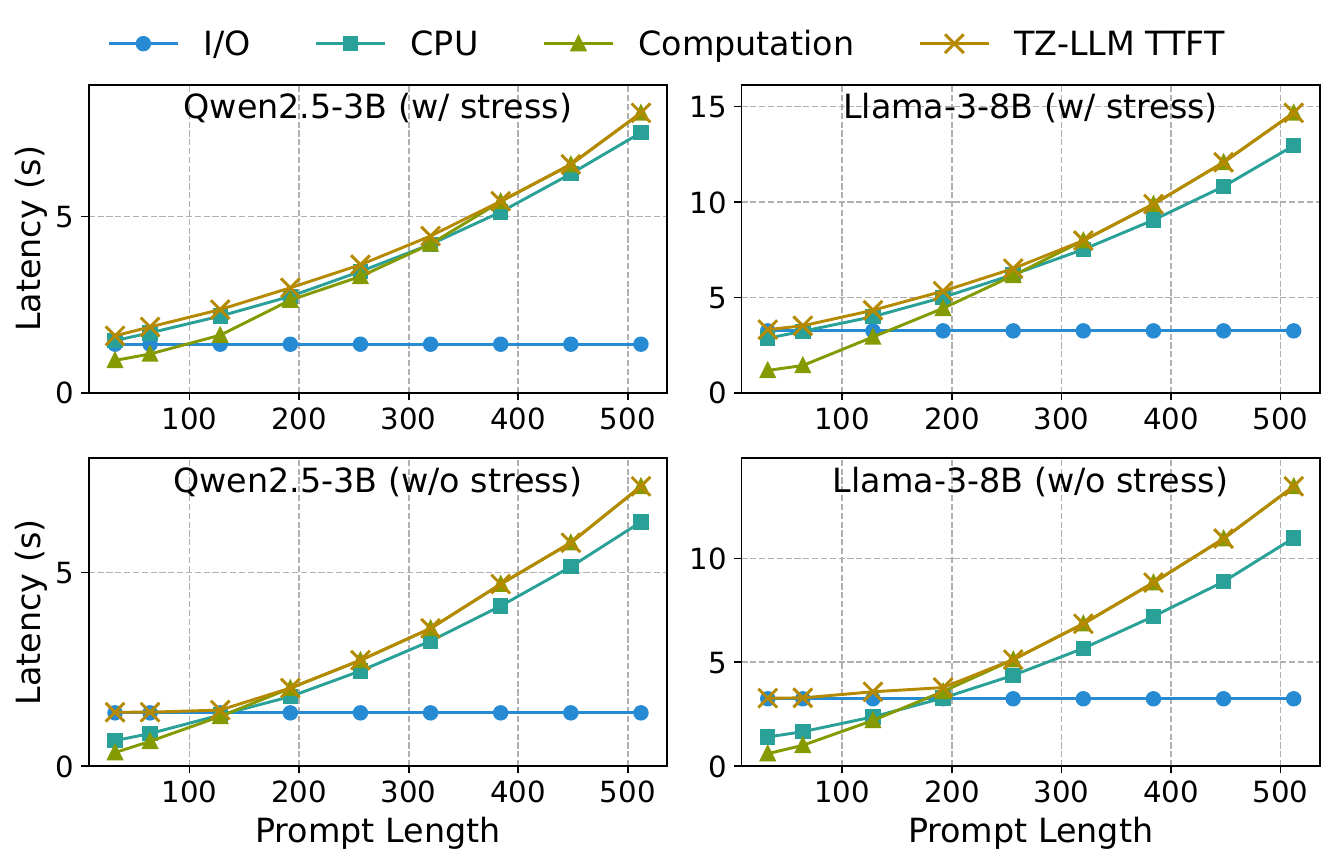}
    \caption{
        The latency of each critical path and the TTFT of {\sys} under different models and prompt lengths, with 20\% LLM parameters cached.
        stress: memory stress.
        I/O: the total latency of all loading (I/O) operators.
        CPU: the total latency of CPU computation, allocation, and decryption.
        Computation: the total latency of CPU and NPU computation.
    }
    \label{fig:schedule}
\end{figure}

\subsubsection{Scheduling Policy Effectiveness}
\label{sec:scheduling-policy-effectiveness}

To evaluate the effectiveness of our priority-based pipeline scheduling policy,
we analyze the latency of the three potential critical paths of the pipeline mentioned in \textsection{\ref{sec:pipeline}}.
The maximum latency of them is the theoretical lower bound on TTFT for any scheduling policy.
We configure the experiments with 20\% of the parameters cached to eliminate initial pipeline bubbles,
which is independent of the scheduling policy.
Since our scheduling policy favors the scenario with the critical path of CPU or computation operators,
we also evaluate the scenario with the critical path of I/O operators by eliminating memory stress (eliminating CPU memory migration overhead) to analyze the worst case of our policy.

Figure~\ref{fig:schedule} shows that {\sys} incurs 0.01\%$\sim$9.9\% overhead compared to the theoretical lower bound when memory stress is enabled.
When disabling memory stress, the overhead increases to a modest 10.4\%.
Therefore, our scheduling policy performs close to the optimal one.

\subsubsection{Effect of Preemptive Scheduling}
\label{sec:pipeline-breakdown}

\begin{figure}[htbp]
    \centering
    \includegraphics[width=1.0\columnwidth]{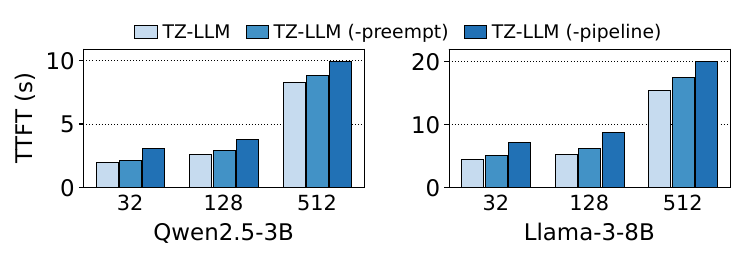}
    \caption{The effect of preemptive pipeline scheduling under different prompt lengths and models.}
    \label{fig:ablation}
\end{figure}

Figure~\ref{fig:ablation} shows the effect of preemptive pipeline scheduling on reducing the TTFT.
Compared with {\sys} without pipelined restoration, the pipeline without preemption reduces the TTFT by up to 31.7\%.
By enabling preemption on allocation and decryption operators, {\sys} eliminates the pipeline bubbles caused by misalignment of operator execution times, further reducing the TTFT by up to 16.2\%.

\subsubsection{Effect of Partial Parameter Caching}
\label{sec:cached-llm-states}

\begin{figure}[htbp]
    \centering
    \includegraphics[width=1.0\columnwidth]{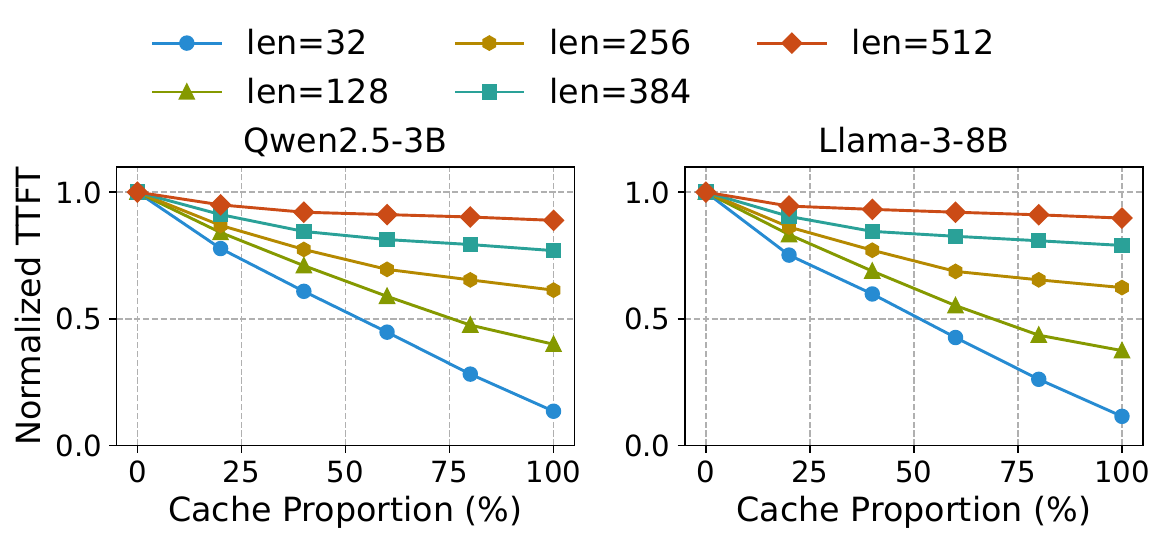}
    \caption{The TTFT of {\sys} under different cache proportions. For each model and prompt length, the TTFT is normalized by the TTFT of the 0\% cache setup.}
    \label{fig:cache}
\end{figure}

To assess the effect of partial parameter caching, we vary the proportion of cached parameters from 0\% to 100\%.
Figure~\ref{fig:cache} illustrates the TTFT of various models across different prompt lengths and cache proportions.

As more parameters are cached, the TTFT decreases approximately linearly up to a threshold.
After this threshold, the benefit of additional caching diminishes as the restoration overhead is effectively hidden beneath the computation.
This threshold is primarily determined by the NPU computation time, which depends on the model and prompt length.
Besides the current mechanism that adjusts the cache size based on REE memory pressure,
{\sys} can also determine a cache size by identifying the threshold with profiling.

\subsection{NPU Time-Sharing Performance}
\label{sec:npu-multiplexing}

\begin{figure}[htbp]
    \centering
    \includegraphics[width=1.0\columnwidth]{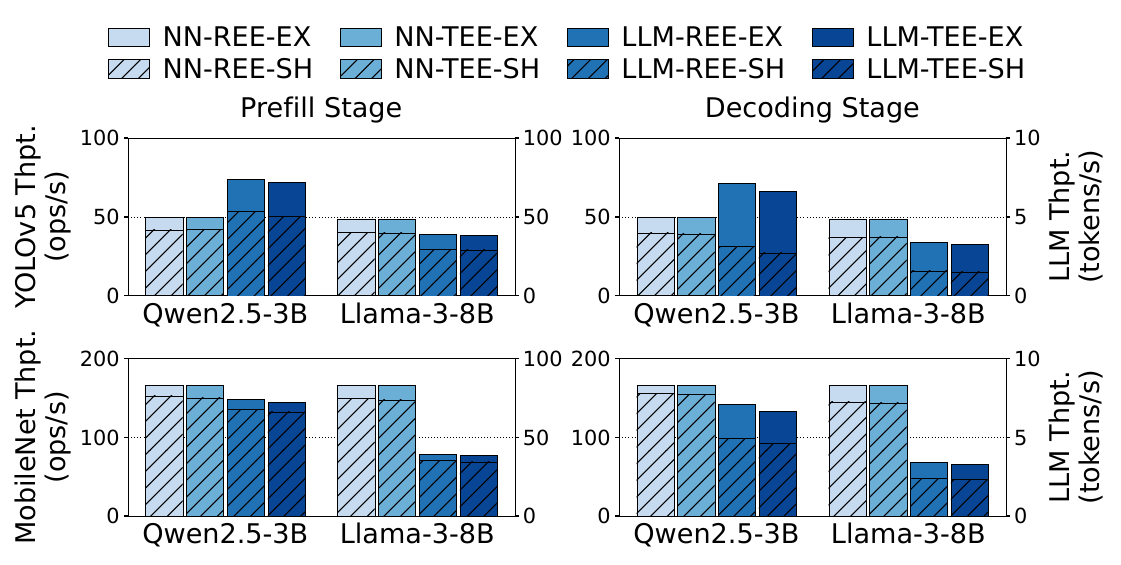}
    \caption{
        The throughputs of NN applications (left y-axis) and LLMs (right y-axis) with NPU time-sharing.
        REE: REE-LLM-Memory, TEE: {\sys} (100\% cached),
        EX: NN application and LLM run exclusively, SH: NN application and LLM run concurrently with a shared NPU.
    }
    \label{fig:npu_multiplexing}
\end{figure}

We evaluate the NPU time-sharing performance of {\sys} by concurrently running mainstream neural network (NN) applications that use the NPU alongside LLM inference.
The two evaluated NN applications are YOLOv5~\cite{yolo_v5} for object detection and MobileNet~\cite{mobilenet} for image classification.
We choose two LLM models with small or large model sizes and use a prompt with 512 tokens.
The throughputs of the NN applications and the LLMs are displayed in Figure~\ref{fig:npu_multiplexing}.

As expected, when the NN application and the LLM run concurrently (-SH), the throughputs of both sides are lower compared with their counterparts under exclusive running (-EX), due to NPU multiplexing.
Compared with NPU time-sharing within the REE (-REE),
the TEE-REE NPU time-sharing mechanism (-TEE) introduces a small additional overhead,
with NN applications and LLMs experiencing only up to 3.8\% and 3.0\% extra slowdown, respectively.

To quantify the overhead of TEE-REE NPU time-sharing, we measure the time spent on (1) {\smc} switches for shadow job scheduling, (2) TZASC and TZPC configuration, and (3) GIC configuration.
The total time-sharing overhead accounts for 1.6\%$\sim$2.7\% and 2.3\%$\sim$5.7\% of the TTFT and decoding time across all evaluated setups.

\subsection{Interference Between CMA and REE Applications}
\label{sec:normal-app-overhead}

\begin{figure}[htbp]
    \centering
    \includegraphics[width=1.0\columnwidth]{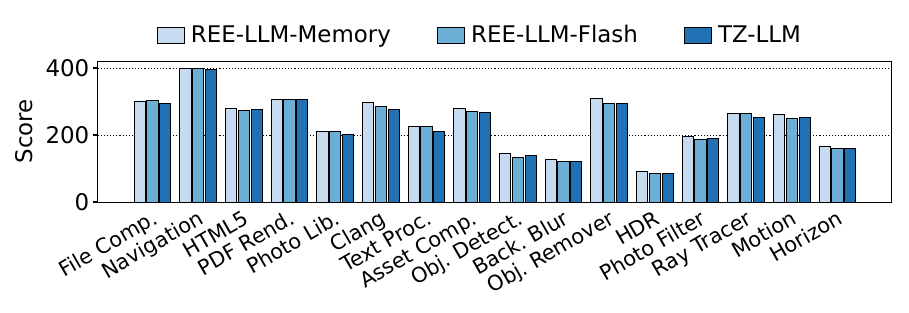}
    \caption{Geekbench scores when running concurrently with LLM prefill stage (Llama-3-8B, 512-token prompt).}
    \label{fig:geekbench}
\end{figure}

The performance of REE applications may be affected by memory migration during CMA allocation.
We evaluate this overhead by concurrently running Geekbench~\cite{geekbench} with LLM inference.
To show the worst-case overhead, we configure {\sys} and REE-LLM-Flash to run the prefill stage, revoke all memory, and then restart the prefill stage.
The REE-LLM-Memory baseline only repeats the prefill stage.
The benchmark threads and LLM inference threads are pinned to different CPU cores.
The results are shown in Figure~\ref{fig:geekbench}.

Compared to the REE-LLM-Memory and REE-LLM-Flash baselines, the Geekbench scores under {\sys} show a degradation of up to 6.7\% and 5.8\%, respectively.
This overhead is comparable to that introduced by S2PT (Figure~\ref{fig:meta-geekbench}).
However, unlike S2PT, the overhead from CMA allocation occurs \emph{only during the prefill stage} and is negligible during the decoding stage or when inference is not running.

%% file: related-work.tex
\section{Other Related Work}
\label{sec:related-work}

\stitle{On-device LLM inference.}
On-device LLM inference has garnered significant attention recently~\cite{on-device-llm,survey-resource-efficient-llm}.
Some prior work runs the LLM with limited memory by swapping parameters between the memory and the flash during inference~\cite{llm-flash,powerinfer-2}.
Combining {\sys} with these parameter offloading techniques is considered future work.
Currently, {\sys} keeps parameters in memory during decoding.
Some prior systems~\cite{powerinfer-2,mllm} use on-device NPUs to speed up LLMs.
{\sys} supports NPU acceleration in the TEE.

Some work reduces the parameter sizes of LLMs with quantization, knowledge distillation, or weight pruning~\cite{gptq,k-quants,awq,zeroquant,gpt3-int8,1-bit-llm,mini-llm,coco-pie,llm-pruner,sparsegpt}.
However, even with reduced parameter sizes, dynamic secure memory scaling is still needed for memory efficiency.

\stitle{Model confidentiality protection.}
Some prior work protects models with cryptographic techniques like homomorphic encryption~\cite{gazelle,delphi} and multi-party computation~\cite{mpc-inference,puma}.
However, they incur significant performance degradation, particularly on resource-constrained mobile devices.
Some systems~\cite{pipellm,10643934} run LLM inference in confidential virtual machines (CVMs)~\cite{intel-tdx,amd-sev,arm-cca,nvidia-cc},
but CVM hardware is not currently available on mobile devices.

\stitle{Outsource-and-verify principle.}
Prior work~\cite{zhou2014dancing} applies the outsource-and-verify principle to the USB driver,
delegating complex USB bus functions to the untrusted OS while verifying their results.
{\sys} adopts this design principle and addresses the specific challenges of the NPU driver,
namely partitioning it into isolated domains and enabling efficient interaction between them.
On the one hand, by examining the workflow of the NPU driver, {\sys} partitions the driver into two isolated domains: the control plane and the data plane.
On the other hand, {\sys} develops efficient methods for the trusted data plane to verify the outcomes of the untrusted control plane.

\stitle{TEEs with accelerators.}
Some prior work extends TEE protection boundary to accelerators in the cloud~\cite{nvidia-cc,telekine,graviton,hix,gevisor,cage,acai,hetee,honeycomb,cure,cronus} or on end devices~\cite{secdeep,strongbox,snpu}.
StrongBox~\cite{strongbox} utilizes the EL3 monitor to protect secure GPU jobs,
which extends the highly privileged TCB.
{\sys} supports NPU in TEE with minimal security impact.
sNPU~\cite{snpu} supports fine-grained and dynamic TEE-REE NPU space-sharing, but it requires hardware modifications.

%% file: conclusion.tex
\section{Conclusion}
\label{sec:conclusion}

{\sys} is a novel system designed for protecting on-device LLMs with Arm TrustZone.
It satisfies LLM performance, memory efficiency, and security requirements using
elastic secure memory scaling with pipelined restoration and TEE-REE NPU time-sharing with control-data separation.

%% file: appendix.tex
\appendix

\section{Artifact Appendix}

\subsection{Abstract}
The artifact contains the source code of our prototype system and the scripts for conducting the experiments presented in the paper.

\subsection{Description \& Requirements}

The artifact is available at \url{https://doi.org/10.5281/zenodo.17213486}.

\subsubsection{Hardware Dependencies}
The artifact requires an Orange Pi 5 Plus board~\cite{orange-pi} (RK3588 CPU/NPU~\cite{rk3588}).
A standalone machine is required to build the artifact and communicate with the board using a USB-to-USB cable.
The machine should have at least 8 GB of memory and 100 GB of free disk space.

\subsubsection{Software Dependencies}
The standalone machine should operate on a Linux OS (Ubuntu 22.04.3 tested)
and have the following dependencies installed:
OpenHarmony Device Connector for board communication,
Python3 and its matplotlib library for plotting and analysis,
and Docker for containerized builds.

\subsection{Setup}
The user needs to download the source code.
Please refer to the \texttt{README.md} for details.

\subsection{Evaluation Workflow}

\subsubsection{Major Claims}

\begin{itemize}
    \item Claim \textbf{C1}:
        {\sys} can reduce TTFT by 76.1\%$\sim$90.9\% compared to the Strawman baseline
        and incurs 5.2\%$\sim$28.3\% overhead compared to the REE-LLM-Memory baseline.
        This is demonstrated by experiment \textbf{E1}, whose results are reported in Figure~\ref{fig:end2end_prefill_benchmark}.
    \item Claim \textbf{C2}:
        {\sys} can increase decoding speed by 0.9\%$\sim$23.2\% compared to the Strawman baseline
        and incurs 1.3\%$\sim$4.9\% overhead compared to the REE-LLM baseline.
        This is demonstrated by experiment \textbf{E2}, whose results are reported in Figure~\ref{fig:end2end_decoding}.
    \item Claim \textbf{C3}:
        As more parameters are cached, partial parameter caching can reduce TTFT approximately linearly up to a threshold.
        This is demonstrated by experiment \textbf{E3}, whose results are reported in Figure~\ref{fig:cache}.
\end{itemize}

\subsubsection{Experiments}

\begin{itemize}
    \item Experiment \textbf{E1} (approximately 60 compute-minutes):
        Run script \texttt{scripts/1-end-to-end-prefill.sh},
        which evaluates the TTFT of {\sys} and other baselines across different benchmarks.
        The results are displayed in \texttt{plots/figure10.pdf}.
    \item Experiment \textbf{E2} (approximately 20 compute-minutes):
        Run script \texttt{scripts/2-end-to-end-decoding.sh},
        which evaluates the decoding speed of {\sys} and other baselines across different models.
        The results are displayed in \texttt{plots/figure11.pdf}.
    \item Experiment \textbf{E3} (approximately 60 compute-minutes):
        Run script \texttt{scripts/3-caching.sh},
        which evaluates the effect of partial parameter caching on the TTFT of {\sys} across different cache proportions and prompt lengths.
        The results are displayed in \texttt{plots/figure14.pdf}.
\end{itemize}